# Regimes of Coherent Intermittency in the Next Quantum Jump of a Multilevel System

Massimo Porrati[1], Nicholas Penthorn[2], John Rooney[2], Hong Wen Jiang[2], and Seth Putterman[2]

[1]*Center for Cosmology and Particle Physics, Department of Physics, NYU, 726 Broadway, NY NY 10003*

[2]*Department of Physics and Astronomy, UCLA, Los Angeles, CA 90095*

An externally driven quantum system will execute deterministic variations in the occupation of its levels |i> that are interrupted by stochastically spaced quantum jumps. The ability to measure the time between successive quantum jumps provides access to rich multi-time scale behavior [1,2]. Consider for example a 3 level atom with ground state |0>, short lifetime, $\tau_1$ [bright] state |1> and long lifetime, $\tau_2$ [dark] state |2> where the |2> - |1> transition is excluded. [Figure 1].

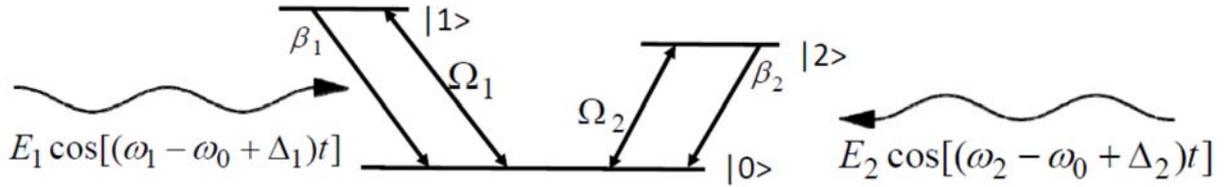

Figure 1. Three level atom where all transitions take place between the excited states |1>, |2> and the ground state |0>. The externally imposed fields $E_1$, $E_2$ can be slightly detuned from the strong |0> - |1> and forbidden |0> - |2> transitions by an amount $\Delta_1$, $\Delta_2$. The Rabi flopping frequencies are $\Omega_1$ and $\Omega_2$ and the spontaneous decay rates are $\beta_1$ and $\beta_2$.

If this atom is excited with lasers tuned to the |2>-|0> and |1>-|0> transitions the coherent evolution of the wave function between photon emissions will vary on time scales due to the Rabi flopping frequencies $[\Omega_1; \Omega_2]$ and lifetimes $\tau_1, \tau_2$. For an isolated ion $\tau_2 \sim 10^8 \tau_1$. The time between successive photon emissions will have intervals determined by *both* $\tau_1, \tau_2$ even in the limit when the exciting field is perfectly coherent [3]. The observation of dark periods of length $\tau_2 >>> \tau_1$ is surprising if one interprets the absence of photon emission as indicating that the atom is deterministically oscillating between its internal states. Such deterministic motion will always have a substantial probability to be in the bright level |1> and therefore only a tiny chance of being dark for a time comparable to $\tau_2$. The

existence of a dark state, in a driven atom, that coherently and deterministically evolves on the long time scale $\tau_2$ is a consequence of being able to observe the emission of the next photon. An observation of the absence of emission may be thought of as weak measurement that smoothly changes the wave function while maintaining phase coherence.

The quantum theory of the next jump was first formulated for the resonance fluorescence of this 3 levels system which has been experimentally [4] realized in an ion trap. Single ions not only have a long coherence time, $\tau_2 \sim 1.0$ sec., for the evolution of the wave-function but they can also be driven to saturation which occurs when $\Omega_1 \tau_1 \gg 1$. Other physical systems that provide the opportunity for observation of the next quantum jump include coupled transmon Q-bits [5] and single/few electron states that can be controlled in 2D electron gases that exist at a thin Si interface sandwiched between SiGe. These Q bits exhibit quantum states determined by charge, spin and valley quantum numbers [6-10]. Some of these systems [5] have the advantage of being able to measure the next jump with almost 100% efficiency but the physical nature of these systems has so far precluded access to the saturation regime and in fact require $\Omega_1 \tau_1 \ll 1$. In this paper we apply the theory of the next quantum jump, that was developed in [1] to this limit.

Consider the amplitudes $c_i(t)$ which describe the joint amplitude that the quantum system is in state |i> and that no quantum jumps have occurred [i.e. no fluorescent photon has been emitted] for a time t since a reference time t=0 when the atom was in |0>. The 3 level atom [Figure 1] is excited by lasers tuned to the |0>-|1> and |0>-|2> transitions. In the rotating wave approximation:

$$\frac{dc_2}{dt} = -(\beta_2/2)c_2 - \Omega_2 c_0,$$
$$\frac{dc_1}{dt} = -(\beta_1/2)c_1 - \Omega_1 c_0, \qquad (1)$$
$$\frac{dc_0}{dt} = \Omega_1 * c_1 + \Omega_2 * c_2,$$

where $1/\beta_i = \tau_i$ are the spontaneous decay lifetimes and $|\Omega_i|$ are the Rabi flopping frequencies between the states, and $\bar{c}_o$ of [1] has been replaced with $ic_0$ and we have taken $\Delta_{1,2}=0$. The joint probability that the atom is in some state and the next jump has not yet occurred

$$W(t) = \sum_i |c_i^2(t)| \qquad (2)$$

is not conserved as it must decrease in time due to the possibility of spontaneous decay.

In [1] intermittency in fluorescence was calculated in the limit $|\Omega_2|/\beta_1 << 1$ and when the strong transition is saturated $|\Omega_1|/\beta_1 >> 1$. As mentioned we now take the limit $|\Omega_1|/\beta_1 = \varepsilon << 1$; which has been raised in connection with transmon Q-bits[5]. In this limit the eigenvalues $\lambda$ for the time development of the amplitudes for the next jump obey:

$$(x+\frac{1}{2})(x^2+Bx+C)=0, \qquad (3)$$

where: $x = \lambda/\beta_1$

$$B = \frac{2}{\beta_1}\beta_\ell; \quad C = \frac{\Omega_2^2}{\beta_1^2} + 2\varepsilon^2 \frac{\beta_2}{\beta_1} \qquad (4)$$

and we have introduced the slow rate:

$$\beta_\ell = [\frac{\beta_2}{4} + \beta_1\varepsilon^2] \qquad (5)$$

The eigenvalues are:

$$\lambda_1 = -\beta_1/2; \quad \lambda_\pm = -\beta_\ell \pm \sqrt{\beta_\ell^2 - [\Omega_2^2 + 2\varepsilon^2\beta_1\beta_2]} \qquad (6)$$

First consider the case $|\Omega_2|/\beta_1 << 1$; but the motion is underdamped so that

$$|\Omega_2|^2 >> \beta_\ell^2. \qquad (7)$$

For an atom that has been reset to the ground state at t=0 the next photon amplitudes become to leading orders in: $\varepsilon; \beta_\ell/|\Omega_2|$:

$$c_0 = \cos(\Omega_2 t)\}\exp[-(\beta_\ell t)]\},$$

$$c_1 = +2\varepsilon\exp(-\beta_1 t/2) - 2\varepsilon\cos(\Omega_2 t)\exp[-\beta_\ell t], \quad (8)$$

$$c_2 = -\sin(\Omega_2 t)\exp[-\beta_\ell t].$$

Where we now take $\Omega_{1,2}$ real. The probability that after a quantum jump the continuous observation of emission will result in a dark period is almost unity, being unity up to $O(\varepsilon^2)$. This is also the probability of there being no jumps for a time: $t_0' \gtrsim 4/\beta_1$. Such an observation projects out the long time response so that for times $t > t_0'$ [the so-called dark period] the amplitudes obey:

$$c_0 \sim f\{\cos(\Omega_2 t)\exp[-\beta_l(t-t_0')]\},$$
$$c_1 \sim -2\varepsilon f\{\cos(\Omega_2 t)\exp[-\beta_l(t-t_0')]\}, \quad (9)$$
$$c_2 \sim -f\{\sin(\Omega_2 t)\exp[-\beta_l(t-t_0')]\},$$

where: $f = \sqrt{1/(1-4\varepsilon^2)}$. As compared to saturated Rabi flopping, weak excitation leads to a dark period that is characterized by two long time scales $1/\Omega$; $1/\beta_\ell$. During the dark period the maximum occupancy of the strongly coupled level |1> is down by a factor of $\varepsilon^2$ from the weakly coupled level; but the probability per unit time [averaged over a cycle of $\Omega_2$] that the dark period will end with a photon from |1> is equal to the probability that it ends with a photon from the dark state |2> { for purpose of exposition we take $\beta_2 \leq O(\beta_1\varepsilon^2)$}. The behavior here would be identical to a the behavior of an excited 2 level system [|0>-|2>] except that a jump from the weakly occupied level |1> can interrupt the Rabi flopping. The transition from |1>-|0> occurs when the occupation of the ground state is nearly unity, which can be interpreted as a situation in which the ground state is unstable. Figure 2 displays the various time scales for the next quantum jump in this system.

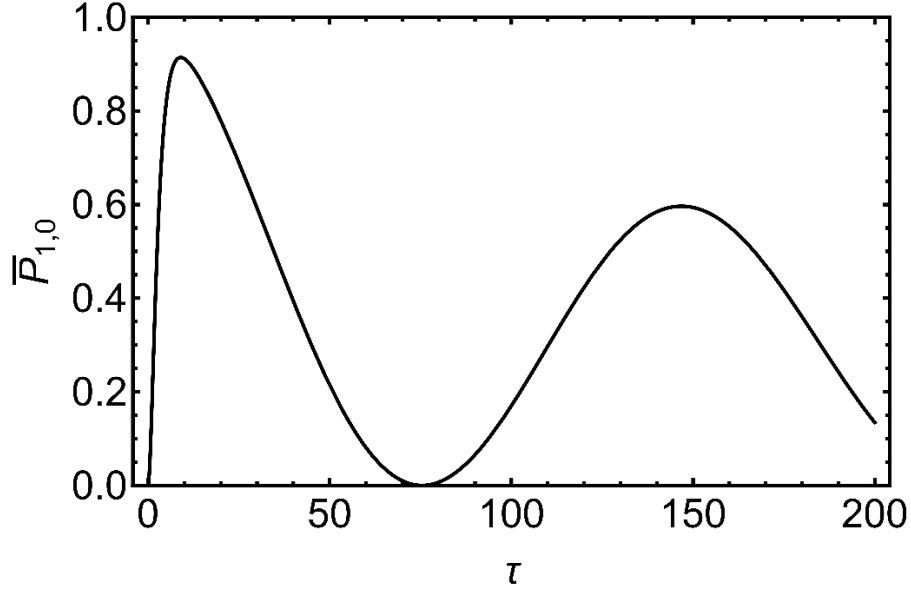

**Figure 2:** Normalized probability $\overline{P}_{1,0} = c_1^2/4\varepsilon^2$ that after a reset to the ground state at t=0, the driven 3 level atom will be in the short lifetime or bright level |1> and that furthermore no photons will have been emitted since the reset. The dimensionless time is $\tau = \beta_1 t$. For this example the Rabi flopping frequencies to the dark and bright levels are $\Omega_2/\beta_1 = 1/48;\ \Omega_1/\beta_1 = \varepsilon = 1/24;\ \beta_2 = 0$. This plot displays 3 time scales for the quantum response. Up until about $\tau$=5 probability to be in |1> rapid builds up; if no emitted photons are observed during this interval the rate of change of the probability becomes over 10 times slower and changes on the time scale determined by $\Omega_2$ The probability that the next photon will be a |1> - |0> photon and that it is emitted in the interval $d\tau$ is $c_1^2 d\tau$: compare Figure 3 of [1]. During the dark period $\tau \gtrsim 5$ [when the motion is dominated by the slow time constants] all three levels are in a coherent superposition. A comparison of this curve to measurements of the distribution of photon emission events will provide a test of the degree of isolation of a quantum system as it evolves.

Next consider the case where $\Omega_2{}^2 \ll \beta_\ell^2$ and as above $\Omega_2/\beta_1 \ll 1$ and $\Omega_1/\beta_1 = \varepsilon \ll 1$. We set $\Omega_2/\Omega_1 = \eta$ and take $\eta^2/4\varepsilon^2 \ll 1$. For the purpose of exposition we also take $\beta_2 \ll \beta_1 \eta^2$. To leading order the eigenvalues are

$$\lambda_1 = -\beta_1/2;\ \lambda_2 = -2\beta_\ell = -2\beta_1\varepsilon^2;\ \lambda_3 = -\eta^2\beta_1/2. \qquad (10)$$

These eigenvalues are real and span 3 well separated time scales.

The initial condition is that $c_0(t \gtrsim 4/\beta_1)=1$, because the weak flopping hasn't had a chance to transfer probability on a time scale $4/\beta_1$, which is approximately when the dark period begins. This also implies that $c_2(t \sim 4/\beta_1) \sim c_{2,2}(0) + c_{2,3}(0) = 0$. The probability that the system will enter a dark period after a reset is again close to unity and during the dark period:

$$c_0 = \exp[\lambda_2 t] - \frac{\eta^2}{4\varepsilon^2}\exp[\lambda_3 t], \qquad (11)$$

$$c_1 = -2\varepsilon \exp[\lambda_2 t] + \frac{\eta^2}{2\varepsilon}\exp[\lambda_3 t],$$

$$c_2 = \frac{\eta}{2\varepsilon}\exp[\lambda_2 t] - \frac{\eta}{2\varepsilon}\exp[\lambda_3 t].$$

The probability that the dark period ends with a photon from the |1> - |0> transition on the time scale $1/|\lambda_2|$ is close to unity. The probability that the next jump occurs on the extra-long time scale $1/|\lambda_3|$ is $\eta^2/4\varepsilon^2$. This is the probability that the system is dark for a time $t_3 > 1/|\lambda_3|$ and when this happens the occupation amplitudes become:

$$c_0 = -\frac{\eta}{2\varepsilon}\exp[\lambda_3 t] \qquad (12)$$

$$c_1 = +\eta \exp[\lambda_3 t]$$

$$c_2 = -\exp[\lambda_3 t]$$

Once the extra-long dark period has been observed the probability that it ends with a photon from the |1>-|0> transition is again close to unity. The probability that the next photon comes from the |2>-|0> transition [during the extra-long dark time] is given by $\beta_2/\beta_1\eta^2$. The probability to be in |2> at long times subject to the null observation of photon emission is shown in Figure 3. Following [5] we consider a measure of the relative occupation of the levels as given by

$Z = [c_2^2(t) - c_0^2(t)] / [c_2^2(t) + c_0^2(t)]$. This quantity goes to -1 at short times where the system is in its ground state and at long times Z asymptotes to $1 - (\eta^2 / 2\varepsilon^2)$. The deviation of Z from unity as $t \to \infty$ is proportional to $\Omega_2^2$ and is therefore essential to the existence of the phenomena under discussion. At an intermediate time: $t_2$ ; Z reaches unity as shown in Figure 3. If the atom is dark for time $t_2$ then it is fully shelved into the dark state |2>. If one verifies that there is no emission for a time $t_2$ then the occupation of the dark level is a maximum and in this approximation the shelving is 100% . Longer dark periods can only reduce certainty about the transition to the dark state as Z never attains the maximum value Z=1 after this time. Furthermore, the complete transfer of probability to |2> has been achieved with exponential as compared to oscillatory processes.

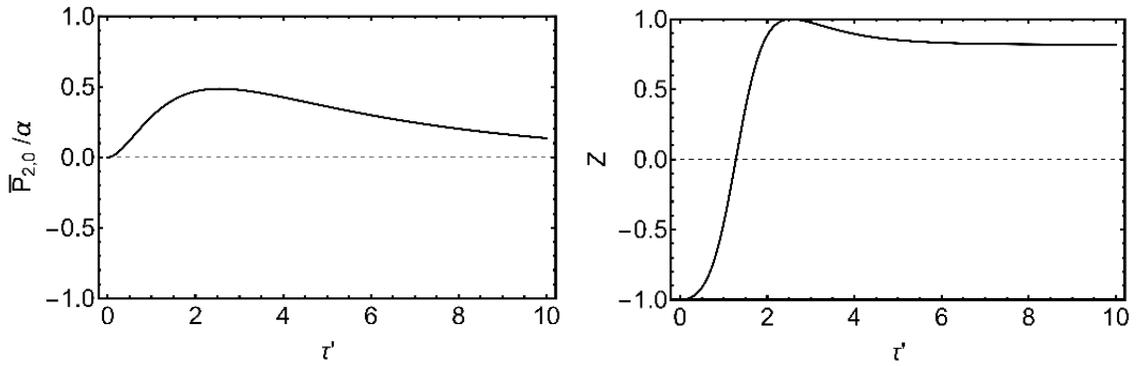

**Figure 3:** Probability P$_{2,0}$ that at time $\tau' = 2\beta_1 \varepsilon^2 t$ the atom will be in |2> and that no photons will have been emitted since the previous reset: $\alpha = (\eta / 2\varepsilon)^2 = 1/10$ for the limit described by Eq 11-(left panel). The relative occupation Z [right panel] shows that during a dark period $c_2$ reaches 100% at a finite time.

In Figure 4 is displayed the response of a 3 level atom with parameters that have been chosen to approximate the values in reference [5].

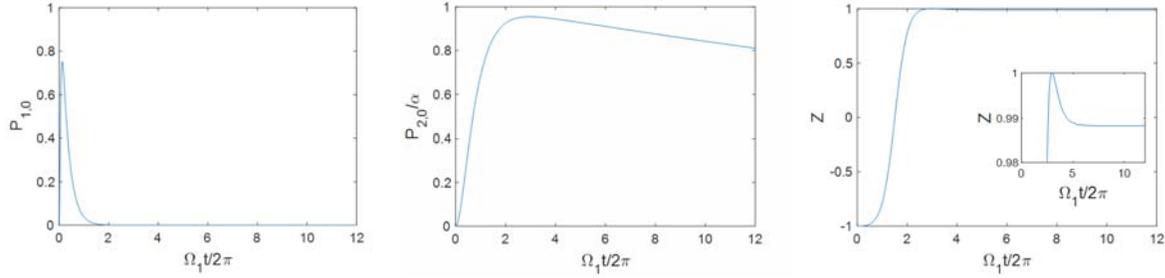

**Figure 4:** Probability that after a reset to the ground state there is no emission until time $t$ and the system will be at this time in $|1\rangle$ [left panel] ; $|2\rangle$ [middle panel] with relative occupation Z [right panel]. For these plots:
$$\Omega_1/2\pi = 1MHz; \Omega_2/2\pi = 20kHz; \beta_1 = 48MHz; \varepsilon = .13; \eta = .02$$

The coherent evolution of a wave function between jumps when ε=1; so that the spontaneous decay rate equals the Rabi flopping rate of the bright level is displayed in Figure 5.

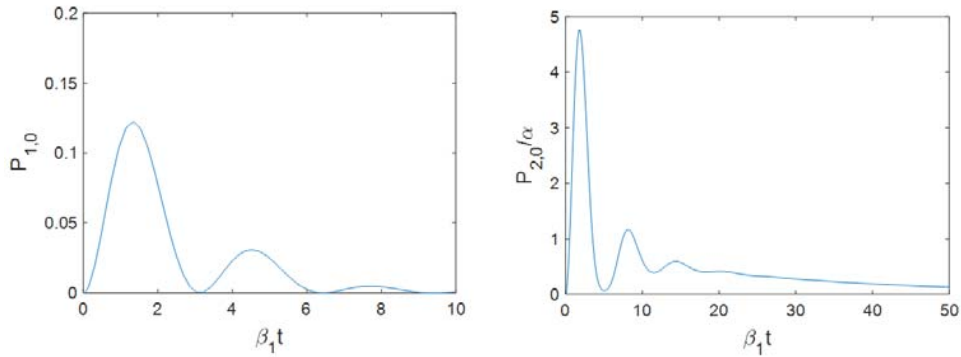

**Figure 5**: Excited state occupation between quantum jumps in the crossover regime where ε=1 : $\Omega_1 = \beta_1 = 10^6 s^{-1}; \Omega_2/\Omega_1 = .2$ .

We have shown that the ability to measure the next photon emission from a 3 level system uncovers time scales much longer than the shortest lifetime [$1/\beta_1$ ]. In the regime of parameter space where the frequency of Rabi flopping is slow compared to the spontaneous decay time of the bright level: $\Omega_1/\beta_1 = \varepsilon << 1$, emission of the next photon is characterized by 3 time scales. When Rabi flopping to the dark level $|2\rangle$ is sufficiently large $\Omega_2^2 >> \beta_\ell^2$ there is an oscillatory probability for the next photon to be recorded from the $|2\rangle$-$|0\rangle$ transition with however a 50:50 probability that the Rabi flopping from $|0\rangle$-$|2\rangle$ is interrupted and reset by an emission from |1>-|0>. When $\eta^2/4\varepsilon^2 = p \ll 1$ there is a probability "$p$" that this driven 3 level system is dark for an extremely long time $t_3 = 2/\eta^2\beta_1$. During this time the atom has an almost 100% probability to be in the dark level |2> yet the transition which ends this extra-long dark period is from state |1>. These behaviors are due to the insight that observation of null emission; namely observation of dark periods, or intervals between quantum jumps, does not alter our ability to describe the atom as a coherently evolving wave function. Therefore we propose that the observation of the statistics of the next quantum jump provided a metric for the quality, and degree of isolation from the environment, of q-bits.

Acknowledgments:

SP wishes to acknowledge valuable discussions with Seth L. Pree

MP is supported in part by NSF grant PHY 1620039

___